\documentclass[twocolumn,prl]{revtex4}
\usepackage{graphics,graphicx}
\begin{document}
\title {Interplay of spin gap and nanoscopic quantum fluctuations in intermetallic CeCu$_{2}$Ge$_{2}$}
\author{D. K.~Singh$^{1,*}$}
\affiliation{$^{1}$NIST Center for Neutron Research, Gaithersburg, MD 20899, USA}
\affiliation{$^{*}$Present address: Department of Physics, University of Missouri, Columbia, MO 65211,USA}

\begin{abstract}
Neutron scattering measurements are used to elucidate interplay of the field-induced spin gap and the quantum fluctuations in archetypal heavy electron system CeCu$_{2}$Ge$_{2}$. A spin gap of $\Delta$ = 0.56 meV is found to develop near the field-induced quantum critical state at $H$$_{c}$$\simeq$ 8 T, driven by fluctuations of the long range antiferromagnetic order parameter. The superconducting transition temperature in many magnetic superconductors is found to exhibit one-to-one correspondence with the spin gap. In this regard, the demonstration of an interplay of the spin gap with the nanoscopic dynamic correlation provides new arena to explore direct relation between the quantum critical behavior and the unconventional superconductivity, especially in heavy electrons systems.
\end{abstract}
\maketitle

The magnetic quantum critical phenomenon (QCP) is of strong importance to correlated electrons systems, as one school of thought believes that occurrence of magnetic QCP is at the core of the unconventional superconductivity in magnetic superconductors.\cite{Fisk,Broun} Intermetallic rare-earth compounds containing a lattice of 4$f$- or 5$f$- electrons (such as Ce, U, and Yb) are prototypical systems to study the magnetic quantum phase transition (QPT)\cite{Maple,Varma}, where the competing interaction between the single-ion Kondo effect and the long range Rudermann-Kittel-Kasuya-Yoshida (RKKY) exchange interaction leads to a novel quantum critical ground state at $T$ $\rightarrow$ 0 K.\cite{Coleman0,Sachdev,Si} Since the Kondo and the RKKY interactions reflect $^{´}$local$^{´}$ and the $^{´}$long range antiferromagnetic$^{´}$ behaviors, respectively, of a system, microscopic nature of the critical fluctuations in the quantum critical state can be dominated by either the local fluctuations or fluctuation of the antiferromagnetic order parameter.\cite{Schroder,Custers}

The standard model of QPT based on the Hertz-Millis-Moriya spin-fluctuation theory invokes that a QPT results from fluctuation of the antiferromagnetic moment (order parameter) and intensity of the order parameter diverges at the QPT.\cite{Hertz,Millis,Moriya} The dynamic behavior, manifested by the spin fluctuations, is experimentally probed using an external control parameter such as pressure, chemical doping or, magnetic field.\cite{Steglich} While synergistic efforts of theoretical and experimental investigations of the quantum critical phenomenon in candidate materials have led to a broader understanding of the underlying mechanism, there are important open questions, however, that need to be addressed in order to develop a universal formulation. One such question is related to understanding the interplay of the quantum fluctuations and the spin gap, which often accompanies a quantum critical state.\cite{Harris,Werlang,Ruegg} It is not clear if occurrence of the spin gap is indeed related to the quantum criticality, especially in some heavy electrons superconductors, such as CeCoIn$_{5}$,\cite{Kenzelmann,Stock}, where the superconducting transition temperature is found to exhibit one-to-one correspondence with the spin gap. Here I report detailed experimental investigation of the field-induced quantum critical behavior and its interplay with the spin gap in stoichiometric CeCu$_{2}$Ge$_{2}$. In a notable observation, it is found that the spin gap, which is absent at zero field, gradually develops to its maximum value of $\Delta$ = 0.56(0.1) meV as applied field is increased to the critical value of $H$$_{c}$$\simeq$8 T. Moreover, the quantum critical state, depicted by the nanoscopic quantum fluctuations, is found to be well described by the HMM spin fluctuation theory.

The compound CeCu$_{2}$Ge$_{2}$ is an unconventional superconductor, $T$$_{c}$$\simeq$0.64 K, under a pressure application of $\simeq$ 10 GPa.\cite{Jaccard} It belongs to Ce$X$$_{2}$$T$$_{2}$ group, where $X$ is a transition metal element and $T$ is $Si$ or $Ge$. Ce$X$$_{2}$$T$$_{2}$ crystallizes in  ThCr$_{2}$Si$_{2}$-type tetragonal structure (I4/mmm space group), where the Ce valence spans the range from a fully trivalent to a strongly mixed valent as $X$ and $T$ are varied.\cite{Grier,Lawrence} Recent researches on single crystal specimens of Ce$X$$_{2}$T$_{2}$ compounds, where $X$ is a nobel metal, have revealed the presence of an interesting near-universal ground state characteristic of spin-density wave configuration at low temperature.\cite{Stockert2,Singh1,Singh2,Singh3} In the archetypal antiferromagnetic metal CeCu$_{2}$Ge$_{2}$, it was concluded that correlated  Ce$^{3+}$ ions form a long range spiral spin density wave ground state configuration below $T$$_{N}$ $\simeq$ 4 K.\cite{Knopp,Singh1} The spiral spin density wave structure, Figure 1d, is illustrated by the temperature dependent incommensurate magnetic positions given by $\textbf{Q}$$_{M}$ = $\tau$$\pm$$\textbf{k}$, where $\tau$ and $\textbf{k}$ are nuclear and modulation vectors, respectively. The periodicity of the amplitude modulation is given by the modulation vector $\textbf{k}$ = (0.285,0.285,0.54) with respect to a simple commensurate structure. In order to understand the evolution of the quantum fluctuations as functions of temperature and magnetic field, detailed neutron scattering measurements are performed on high quality single crystals, grown by the flux method. Characterization of the crystals were performed using X-ray measurements, confirming the high quality of the crystals. The neutron scattering measurements were performed on one rectangular shape single crystal with a mass of 1.8 gm on SPINS and MACS cold spectrometers at the NIST Center for Neutron Research with fixed final energy of 3.5 meV and cooled Be filter after the sample. At this fixed final energy, the spectrometers resolution (FWHM) were determined to be $\simeq$ 0.15 and 0.2 meV, respectively. The horizontal collimations were 32$^{'}$-80$^{'}$-Sample-80$^{'}$-120$^{'}$. The sample was mounted in a superconducting vertical field magnet in the [HHL] scattering plane with $a$ = $b$ = 4.345 $\mathring{A}$ and $c$ = 10.94 $\mathring{A}$. Perpendicular field application direction was (-110) direction in this tetragonal system. 

\begin{figure}[tbp]
\centering
\includegraphics[width=12cm]{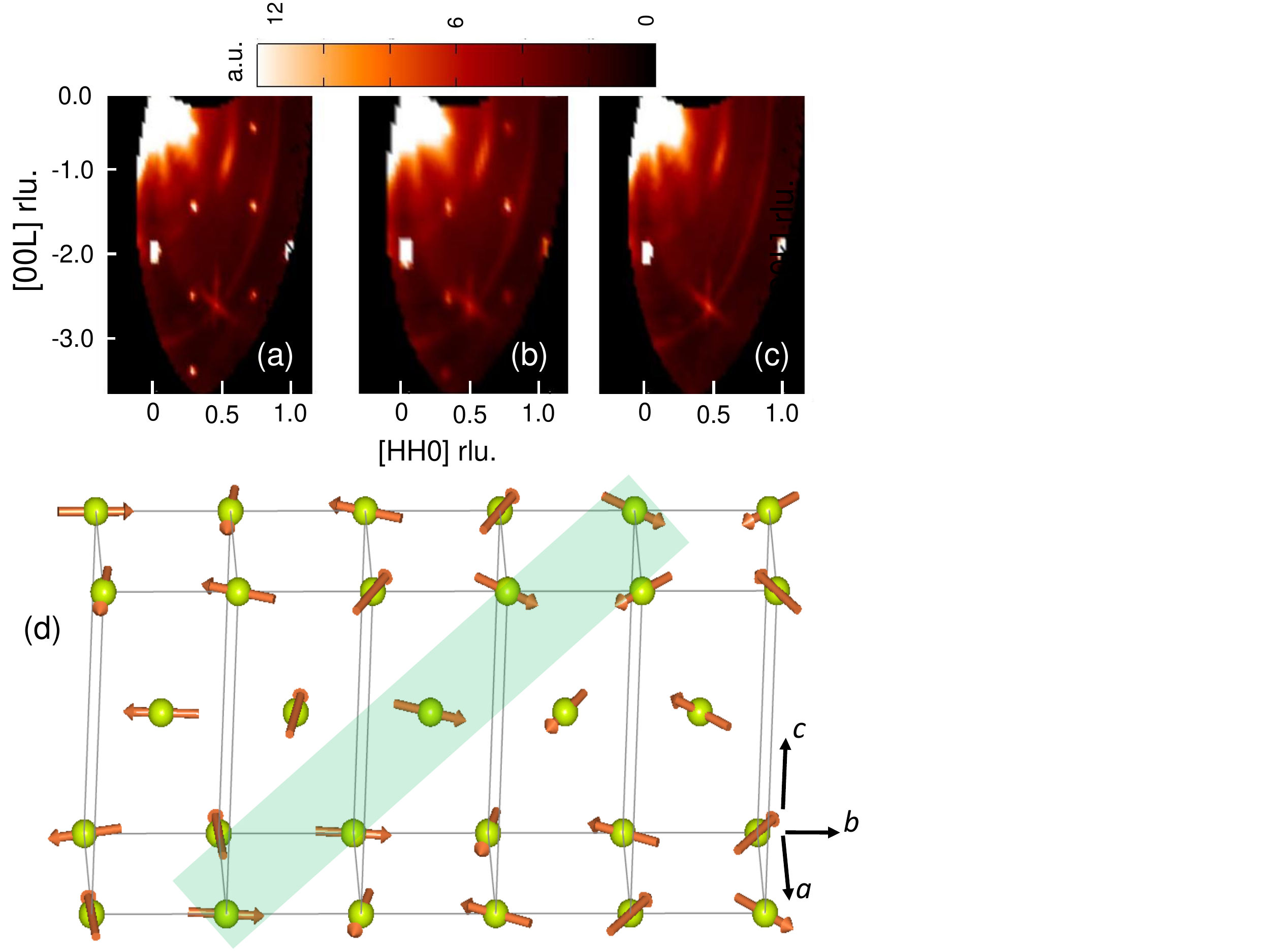} \vspace{-2mm} \vspace{-4mm}
\caption{(Color online) Evolution of elastic map as a function of field and spin structure of correlated Ce ions of CeCu$_{2}$Ge$_{2}$. Elastic maps, measured on MACS at $T$ $\simeq$ 0.2 K, at different fields of (a) $H$ = 0 T, (b) 7 T, and (c) 8 T. Resolution limited magnetic Bragg peaks, small bright dots at incommensurate wave vectors described by the propagation vector $\textbf{k}$ = (0.285,0.285,0.54) with respect to a simple commensurate structure, become weaker as the field is increased and finally disappear at $H$ $\simeq$ 8 T. (d) Correlated Ce moments form a spiral spin density wave configuration  in the $a$-$b$ plane. The shaded area highlights the spiral plane of rotation perpendicular to the propagation vector $\mathbf{k}$. }
\end{figure}

\begin{figure}[tbp]
\centering
\includegraphics[width=13.5cm]{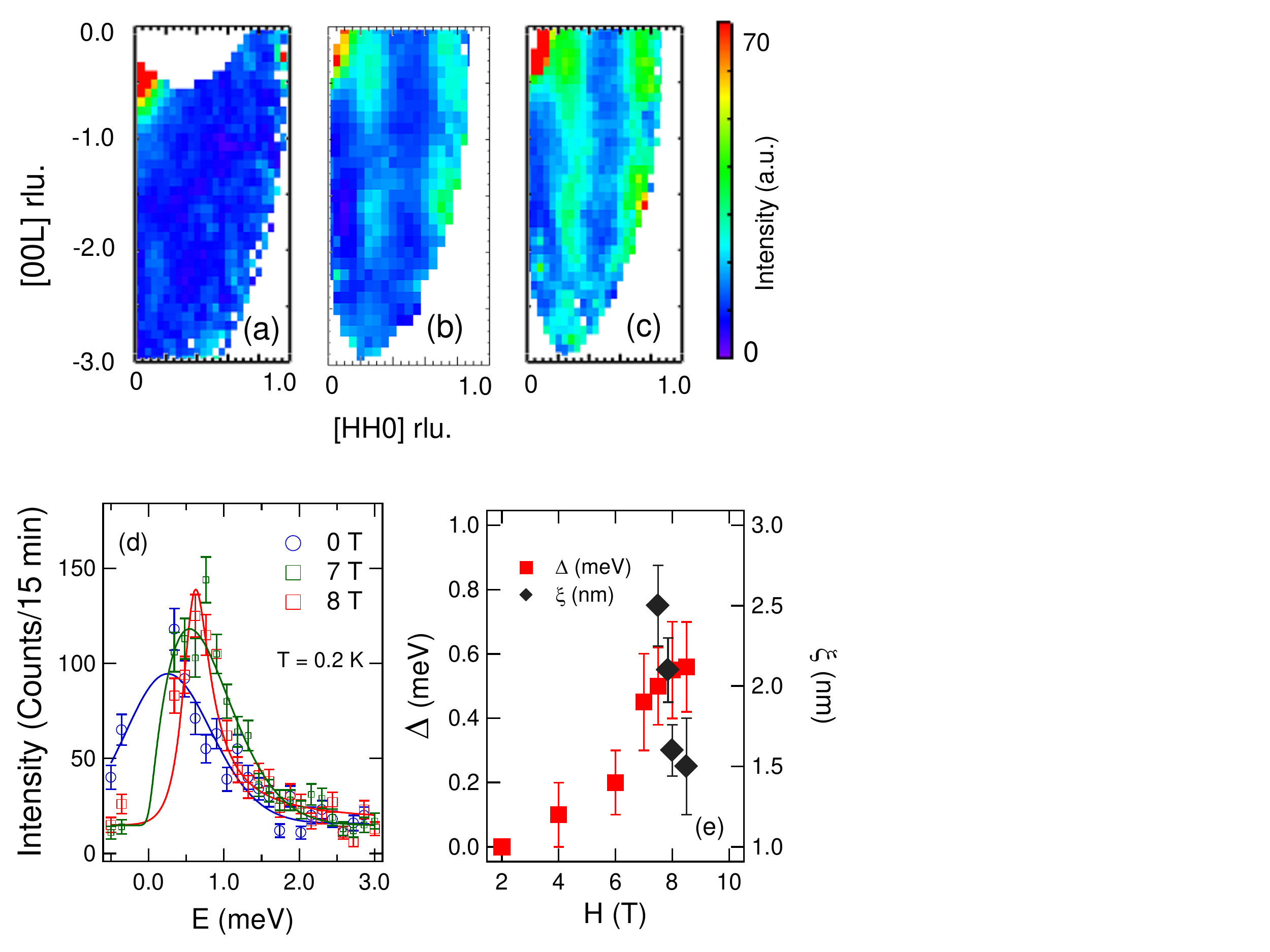} \vspace{-1mm} \vspace{-1mm}
\caption{(Color online) Inelastic neutron scattering measurements as a function of field. (a-c) $q$-scan maps, measured at $T$ $\simeq$0.2 K and at the energy transfer of $E$ = 1 meV, at different fields of $H$ = 7 T (fig. a), 7.5 T (fig. b) and 8 T (fig. c), respectively. At $H$ = 7 T, the dynamic correlation is mostly absent. As the field is increased to the critical value, $H$$_{c}$ = 8 T, short-range dynamic correlation develops at the magnetic wave vector position. Plots are corrected for the aspect ratio between [110]$^{*}$ and [001]$^{*}$. (d) Representative scans of inelastic measurements at the magnetic wave vector, $\textbf{Q}$$_{M}$, at different fields and $T$ $\simeq$ 0.2 K. Solid lines are Lorenztian fits to the data. The system gradually develops a spin gap as the applied field is increased. (e) In order to highlight the interplay between the spin gap and the quantum fluctuation, estimated value of the spin gap, $\Delta$, as a function of field is over-plotted to the correlation length, $\xi$, vs. field plot. }
\end{figure}

A representative contour map of elastic neutron scattering data at $T$ = 0.2 K and $H$ = 0 T, depicting incommensurate magnetic peaks related by the modulation vector $\textbf{k}$ = (0.285,0.285,0.54), is shown in Fig. 1a. The best fit of the experimental data is obtained with the Ce-spins lying within the basal plane and forming a spiral spin density wave with spins pointing in the $a-b$ plane, Fig. 1d. The plane of rotation of the spiral is perpendicular to $\mathbf{k}$.  The static moment associated with the long range incommensurate magnetic peaks, $Q$$_{M}$, at $T$ = 1.5 K was estimated to be M$_{Ce}$ $\simeq $ 1.04(4) $\mu $$_{B}$, significantly smaller than that expected for a fully degenerate $J$ = 5/2 ground state Ce$^{3+}$ ion value of 2.15 $\mu $$_{B}$. The magnitude and direction of the propagation vector is consistent with theoretically determined wave vector connecting parallel planes of the nested Fermi surface of CeCu$_{2}$Ge$_{2}$.\cite{Zwicknagl} In external magnetic field, applied perpendicular to the [HHL] scattering plane, the system exhibits quantum criticality at $H$$_{c}$$\simeq$ 8 T. Detailed measurements are performed as a function of field to understand the development of the quantum critical state. Representative contour maps at various fields at $T$ $\simeq$0.2 K are plotted in Figure 1a-c. A gradual disappearance of the long range AFM order is observed as applied field is increased. The static long range order is replaced by a short range dynamic correlation, which tends to become stronger as $H$ $\rightarrow$ $H$$_{c}$ (Fig. 2a-c); suggesting the quantum-mechanical nature of magnetic instability.\cite{Hewson,Sachdev} The evolution of short-range dynamic correlation is illustrated in Figure 2a-c. The contour maps in Figure 2a-c were measured at different fields at a finite energy transfer of $E$ =  1 meV. At $H$ =  7 T, the system is inside the AFM regime (see Figure 1b). Hence, no dynamic structure factor is observed in Figure 2a. As applied field approaches the critical value, $H$ = 7.5 T, short-range dynamic correlation starts appearing (Figure 2b). The finite size spatial correlation in Figure 2b becomes broader and more intense as the system crosses into the magnetic instability regime at $H$$\simeq$$H$$_{c}$, Figure 2c.

More information about the development of short-range dynamic correlation is obtained by performing quantitative analysis of inelastic measurements. Representative inelastic scans at different fields at $T$ = 0.2 K are plotted in Figure 2d. The background subtracted experimental data is well fitted with instrument resolution convoluted Lorentzian line-shape multiplied by the detailed balance factor. It is immediately noticed that the inelastic spectral line is gapless at $H$ = 0 T. CeCu$_{2}$Ge$_{2}$ gradually develops a spin gap, as evident from energy scans in Figure 2d, and attains an optimum value of $\Delta$ = 0.56(0.1) meV at the critical field. Correlation length of the short-range dynamic struture, $\xi$, is estimated within a finite-size model of the Gaussian linewidth, which fits the two-dimensional cut across $\textbf{Q}$$_{M}$ in false color map (as illustrated in Fig. 2a-c). The plot in Figure 2e illustrates a quantitative interplay between the spin gap, $\Delta$, and the nanoscopic correlation length, $\xi$, of the short-range dynamic structure. As the system approaches QCP, the short-range dynamic correlation length decreases while the spin gap increases. 

The quantum mechanical nature of the nanoscopic fluctuation is established by measuring the temporal fluctuation as a function of temperature. The temporal fluctuation is defined by inverse of the line-width of inelastic spectral line.\cite{Si2} Incidentally, inelastic measurements as a function of temperature can also be used to elucidate the underlying mechanism of the quantum fluctuations, thus associated quantum phase transition.\cite{Knafo,Lohneysen} The characteristic signature of a quantum phase transition is manifested by the determination
of the critical slowing down of the temporal fluctuation (or the relaxation rate) and is reflected in the energy
linewidth of the excitation spectrum.\cite{Coleman0,Sachdev} In the HMM
formalism, the divergence of the static susceptibility as $T$ $\rightarrow $
0 K is also expected.\cite{Knafo} Alternatively, if the quantum critical
behavior is dominated by the $local$ Kondo interaction, the static
susceptibility follows the Curie-Weiss ansatz.\cite{Schroder,Si} In
order to determine the linewidth of the excitation spectrum, detailed measurements were performed as a function of energy at the magnetic wave-vecor $Q$$_{M}$ = (0.29,0.29,-0.54) at various temperatures between $T$ = 0.2 K and
40 K. Figure 3a shows the background corrected representative scans at $Q$$_{M}$ at
few selected temperatures, above and below $T$$_{N}$, and at $H$ $\simeq $
8 T. Plots in Figure 3a confirm the qualitative behavior observed in the
field dependence of $\mathbf{q}$-scans in Figure 2, where a strong
enhancement in the antiferromagnetic fluctuations is observed at $H$$_{c}$. At high temperatures,
the spin fluctuations at the AFM wave vectors cross into the thermal
regime.

\begin{figure}[tbp]
\centering
\includegraphics[width=10cm]{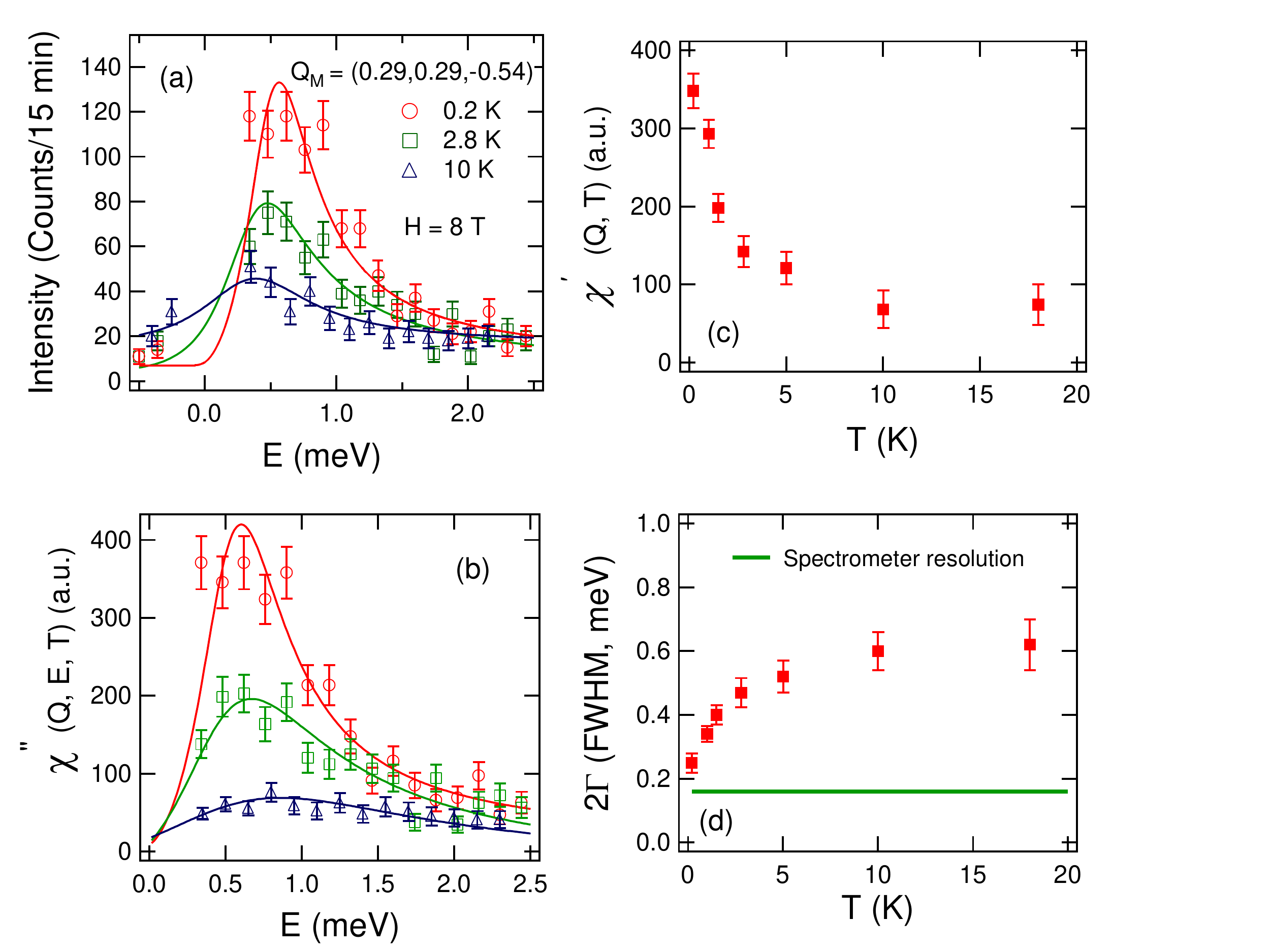} \vspace{-2mm} \vspace{-1mm}
\caption{(Color online) Inelastic measurements as a function of temperature. (a) Representative scans of inelastic measurements at the magnetic wave vector, $\mathbf{Q}$$_{M}$, at different temperatures and $H$ $\simeq$ 8 T. Solid lines are Lorenztian fits to the data, as explained in
the text. Error bars represent one standard deviation. (b) Dynamic
spin susceptibilities at $\mathbf{Q}$$_{M}$ are
extracted from the inelastic neutron scattering data. (c)
Temperature dependence of the static susceptibility at $\mathbf{Q}$$_{M}$
as a function of temperature. The static susceptibility at $\mathbf{Q}$$_{M}$
tends to diverge as $T$$\rightarrow$0 K. (d) Temperature variation
of the relaxation rate, 2$\Gamma(Q, T)$ (meV), as a function of temperature at $H$ = 8 T.
Clearly, the long wavelength spin fluctuations critically slow down at $%
\mathbf{Q}$$_{M}$ as $T$$\rightarrow$0 K. }
\end{figure}

The quantitative determinations of the relaxation rate and the
static susceptibility as a function of temperature are performed using standard
neutron scattering intensity analysis,\cite{Knafo2} which can be
described by,

\begin{eqnarray}
{S(Q, E, T)}&=&\frac{1}{\pi}\frac{1}{1- {e}^{-E/kT}} {\chi}%
^{^{\prime\prime}}(Q, E, T)
\end{eqnarray}

where $\chi$$^{^{\prime\prime}}$ is the dynamic spin susceptibility. The dynamic susceptibility is described by the Fourier transform of the exponential decay of relaxation rate $\Gamma$(Q,T)and is related to the total and static susceptibilities $\chi$ and $\chi$$^{^{\prime}}$, respectively, via the Kramers-Kronig relation. $\chi $$^{^{\prime }}$$(Q,T)$ was determined by integrating $\chi $$^{"}$(Q, E, T)/$E$ over the experimental energy range between 0 and 3 meV, as the energy spectra of fluctuations in this case are limited to low energy.\cite{Knafo2} Inelastic neutron scattering data in Fig. 3a are well fitted using the above equations, giving a single Lorentzian lineshape. The extracted values of $\chi $$^{^{\prime }}$$(Q,T)$ and 2$\Gamma $$(Q,T)$ (full width at half maximum, FWHM) at $Q$$_{M}$ are plotted as a function of temperature in Figure 3c and 3d. In these plots, we see that a continuous increase in the static susceptibility is accompanied by a continuous decrease in the linewidth at $Q$$_{M}$ as the temperature decreases. The decrease in $\Gamma $(Q) becomes faster as the system passes through the zero-field AFM transition at $\simeq $ 4 K and critically slows down as $T$$\rightarrow $0 K. Also noticeable is the diverging behavior in $\chi$$^{'}$$(T)$ as the temperature is reduced to 0.2K. Similar measurements and analysis of inelastic spectra at a non-magnetic wave vector, $\textbf{Q}$$_{0}$ far away from $\textbf{Q}$$_{M}$, did not exhibit any of these behaviors. Rather an almost temperature independent $\Gamma $$(Q,T)$ is observed, which can be used to deduce characteristic temperature of the local Kondo interaction, $T$$^{*}$$\simeq$6.5 K.\cite{Knafo} Parallel presence of the Kondo screening with a localized antiferromagnetic interaction poses an important question: are the bulk properties dominated by the local Kondo screening term or the magnetic wave-vector dependent fluctuations of the AFM order parameter. This argument is further analyzed by performing a scaling analysis of the dynamic spin susceptibility data in the form $\chi $$^{''}$.$T$$^{\alpha }$ = $f$($E$/$T$$^{\beta }$).\cite{Knafo2,Schroder} It is argued that the Curie-Weiss behavior, reflecting the local moment fluctuations, leads to a linear ($E$/$T$) scaling.\cite{Si} On the other hand, if the spins fluctuations are dominated by the AFM order parameter then $\beta$ is 1.5.\cite{Millis,Stockert2,Continentino} As shown in Fig. 4, the scaling plot of dynamic susceptibilities, obtained using equation 1, indeed collapse onto one curve for the SDW-type scenario: $\alpha $ = 1.5, $\beta $ = 1.5.  Experimental data in Fig. 4 is fitted using the HMM formulation leading to $f(x)$ = {abx}/{(1+(bx)$^{2}$)}. The scaling of the dynamic susceptibilities using the HMM formulation further corroborates the view that the field-induced quantum fluctuations in CeCu$_{2}$Ge$_{2}$ are primarily of the long wavelength antiferromagnetic origin.

\begin{figure}[tbp]
\centering
\includegraphics[width=10cm]{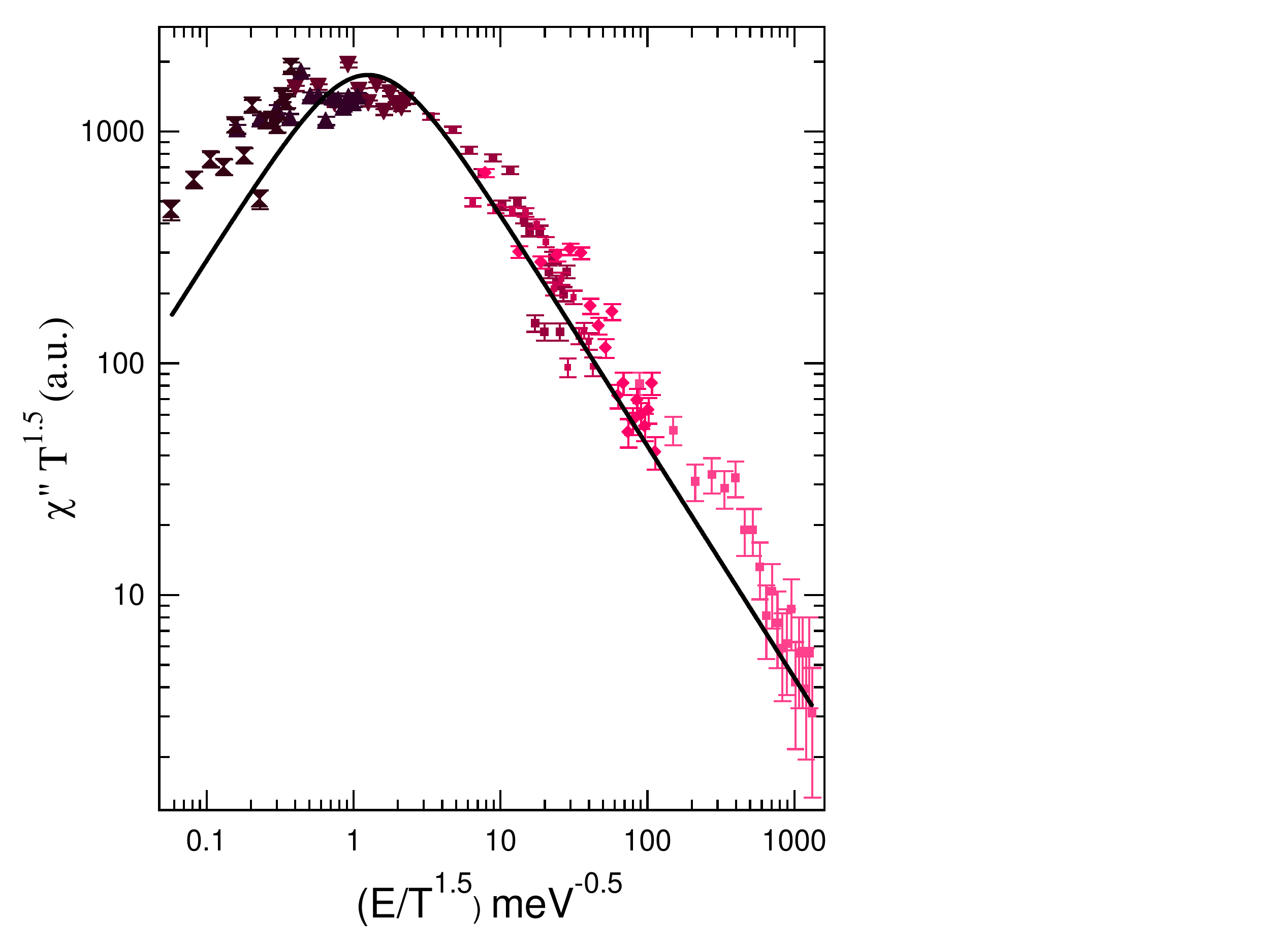} \vspace{-2mm} \vspace{-1mm}
\caption{(Color online) Scaling plot of the dynamic susceptibility data at various temperatures in CeCu$_{2}$Ge$_{2}$. Scaling of dynamic susceptibilities is best described by HMM spin fluctuation theory (see text). }
\end{figure}

Detailed experimental investigation of the stoichiometric compound CeCu$_{2}$Ge$_{2}$ has twofold implications: (1) An important finding is the identification of the fluctuations of the AFM order parameter, well described by the HMM spin fluctuation theory, as the underlying mechanism behind the quantum critical phenomenon. Recently, there has been considerable discussion about the applicability of the HMM formulation in explaining the quantum critical phenomenon.\cite{Coleman,Anderson} This report sheds new light in this regard, as the stoichiometric nature of the system rules out any effect due to chemical disorder. (2) Most important finding, perhaps, is the interplay of the intrinsic spin gap with the quantum phase transition in applied field. Many heavy electron systems exhibit dual phenomena of the quantum phase transition and the unconventional superconductivity. In some cases, the superconducting transition temperature is found to exhibit one-to-one correspondence with the spin gap. Strong evidence of the interplay between the spin gap and the QPT in CeCu$_{2}$Ge$_{2}$ provides new arena to directly relate QPT to occurrence of the unconventional superconductivity. This observation gains more prominence when we consider the fact that majority of the unconventional superocnductors, such as high temperature cuprates or Fe-based pnictides,\cite{Tranquada,Xu} have magnetic origin, where one-to-one correspondence between the spin gap and the superconducting transition temperature is not so uncommon. CeCu$_{2}$Ge$_{2}$ also exhibits unconventional superconductivity at $T$$_{c}$$\simeq$0.64 K under the pressure application of $\simeq$10 GPa. The technical difficulties associated in accessing this pressure regime for neutron scattering measurements makes the understanding incomplete. However, both the spin gap and the QPT are intrinsic properties of a system. Therefore, future analytical works connecting the pressure induced superconductivity to the field induced QPT using energetic analysis will be very helpful.

This work used facilities supported in part by the NSF under Agreement No.
DMR-0944772. I thank A. Thamizhavel, S. K. Dhar, C. Stock, Y. Zhao and S. Chang for their kind help at various stages of the project.


\begin{thebibliography}{99}

\bibitem{Fisk} Z. Fisk \emph{et al.}, \textit{Science} \textbf{239}, 33 (1988).

\bibitem{Broun} D. M. Broun, \textit{Nature Physics} \textbf{4}, 170 (2008).

\bibitem{Maple} M. B. Maple \emph{et al.}, \textit{J. Low Temp. Phys.} \textbf{161}, 4 (2010).

\bibitem{Varma} C. M. Varma, \textit{Rev. Mod. Phys.} \textbf{48}, 219-238 (1976).

\bibitem{Coleman0} P. Coleman and A. Schofield, \textit{Nature} \textbf{433}, 226-229 (2005).

\bibitem{Sachdev} S. Sachdev, Quantum Phase Transitions (Cambridge Univ. Press, New York, 1999).

\bibitem{Si} Q. Si \emph{et al.}, \textit{Nature} \textbf{413}, 804 (2001).

\bibitem{Schroder} A. Schroder \emph{et al.}, \textit{Nature} \textbf{407}, 351 (2000).

\bibitem{Custers} J. Custers \emph{et al.}, \textit{Nature} \textbf{424}, 524 (2003).

\bibitem{Hertz} J. A. Hertz, \textit{Phys. Rev. B} \textbf{14}, 1165 (1976).

\bibitem{Millis} A. J. Millis, \textit{Phys. Rev. B} \textbf{48}, 7183
(1993).

\bibitem{Moriya} T. Moriya \emph{et al.}, \textit{J. Phys. Soc. Jpn} \textbf{64}%
, 960 (1995).

\bibitem{Steglich} P. Gegenwart \emph{et al.}, \textit{Nature Physics} \textbf{4}, 186 (2008).

\bibitem{Harris} A. B. Harris, \textit{J. Mag. Mag. Mat.} \textbf{226-230}, 524 (2001).

\bibitem{Werlang} T. Werlang \emph{et al.}, \textit{Phys. Rev. Lett.} \textbf{105}, 095702 (2010).

\bibitem{Ruegg} Ch. Ruegg \emph{et al.}, \textit{Phys. Rev. Lett.} \textbf{93}, 257201 (2004).

\bibitem{Kenzelmann} M. Kenzelmann \emph{et al.}, \textit{Science} \textbf{321}, 1652 (2008).

\bibitem{Stock} C. Stock \emph{et al.}, \textit{Phys. Rev. Lett.} \textbf{100}, 087001 (2008).

\bibitem{Jaccard} D. Jaccard \emph{et al.}, \textit{Physics Letters A} \textbf{163}, 475 (1992).

\bibitem{Grier} B. H. Grier \emph{et al.}, \textit{Phys. Rev. B} \textbf{29}, 2664 (1984).

\bibitem{Lawrence} J. M. Lawrence \emph{et al.}, \textit{Rep. Prog. Phys.} \textbf{44}, 1 (1981).

\bibitem{Stockert2} O. Stockert \emph{et al.}, \textit{Phys. Rev. Lett.} \textbf{99}, 237203
(2007).

\bibitem{Singh1} D. K. Singh \emph{et al.}, \textit{Scientific Reports} \textbf{1}, 119
(2011).

\bibitem{Singh2} D. K. Singh \emph{et al.}, \textit{Physical Review B} \textbf{84}, 052401 (2011).

\bibitem{Singh3} D. K. Singh \emph{et al.}, \textit{Physical Review B (Rapid)} \textbf{86}, 060405(R) (2012).

\bibitem{Knopp} G. Knopp \emph{et al.}, \textit{Z. Phys. B} \textbf{77}, 95
(1989).

\bibitem{Zwicknagl} G. Zwicknagl, \textit{J. Low Temp. Phys.} \textbf{147}, 123 (2007).

\bibitem{Hewson} A. C. Hewson, The Kondo problem to Heavy Fermions. (Cambridge Univ. Press, Cambridge, 1993).


\bibitem{Si2} Q. Si and F. Steglich, \textit{Science} \textbf{329}, 1161 (2010).

\bibitem{Knafo} W. Knafo \emph{et al.}, \textit{Nature Physics} \textbf{5}, 753
(2009).

\bibitem{Lohneysen} H. Lohneysen \emph{et al.}, \textit{Rev. Mod. Phys.} \textbf{79}, 1015
(2007).

\bibitem{Knafo2} W. Knafo \emph{et al.}, \textit{Phys. Rev. B} \textbf{70}, 174401 (2004).

\bibitem{Continentino} M. A. Continentino, Quantum Scaling in Many-Body Systems (World Scientific, 2001).

\bibitem{Coleman} P. Coleman \emph{et al.}, \textit{J. Phys.: Cond. Matt.} \textbf{13}, R723 (2001).

\bibitem{Anderson} P. W. Anderson \emph{et al.}, Preprint at http://arxiv.org/abs/0810.0279v1 (2008).

\bibitem{Tranquada} J. M. Tranquada \emph{et al.}, \textit{Nature} \textbf{375}, 561 (1995).

\bibitem{Xu} Z. Xu \emph{et al.}, \textit{Phys. Rev. B} \textbf{82}, 104525 (2010).




\end{thebibliography}
\end{document}